\documentclass[12pt]{article}
\usepackage{blois,epsfig}
\begin{document}

\vspace{4mm}

\title{Baryon Production at LHC Energies and Very High Energy Cosmic Ray Spectra.}

\author{Olga I. Piskounova}

\address{P.N.Lebedev Physics Institute of Russian Academy of Science, Moscow, Russia}

\maketitle\abstract{
 The spectra of baryons at LHC can explain the features of the proton spectra in Cosmic Rays (CR). It seems important to study all baryon data that are available from collider experiments in wide range of energies. 
Transverse momentum spectra of baryons from RHIC ($\sqrt(s)$=62 and 200 GeV) and from LHC ($\sqrt(s)$=0.9 and 7 TeV) have been considered. It is seen that the slope of distributions at low $p_T$'s is changing with energy. The QGSM fit of these spectra gives the average transverse momenta which behave as $s^{0.06}$ that is similar to the previously observed behavior of $\Lambda^0$ hyperon spectra. The change in average transverse momenta that are slowly growing in VHE hadron interactions at CR detectors cannot cause the "knee" in measured cosmic ray proton spectra. 
In addition, the available data on heavy quark hadron production from LHC-b at $\sqrt{s}$=7 TeV were also studied. The preliminary dependence of hadron average transverse momenta on their masses at LHC energy is presented.
The possible source of cosmic ray antiparticle-to-particle ratios that are growing with energy was analyzed in the framework of QGSM, where the growing ratios are the result of local leading asymmetry between the production spectra of baryons and antibaryons in the kinematical region of proton target fragmentation. 
In the laboratory system of cosmic ray measurements this spectrum asymmetry will be seen as growing ratio of secondary antiparticle-to-particle spectra until the certain energy of secondary's. This conclusion makes the particle production at the sources of very high energy cosmic protons important, if the interactions with positive target matter would have place in proximity of these sources.
}
\newpage

\section{Introduction}
\label{intro}

The spectra of baryons at LHC can explain the features of cosmic ray particle spectra at very high energies. The transverse momentum distributions are the primary data that can be obtained in the study of hadron spectra at the modern colliders. Interpretation of these distributions in up-to-date phenomenological models can shed a light on the physics of hadroproduction processes at high energies. The phenomenological approach is applied here to the description of $p_T$ spectrum of various sorts of baryons in the framework of Quark-Gluon String Model \cite {QGSM}. 
The model has described the data of previous colliders up to energies $\sqrt(s)= 53 GeV$ at the area of low $p_T$'s that gives main contribution to the average value of transverse momenta \cite {veselov}. Recently $\Lambda^0$ production has been studied \cite{Lambda} in updated version of this model.  

\begin{figure}[htpb]
  \centering
  \includegraphics[width=8.0cm, angle=0]{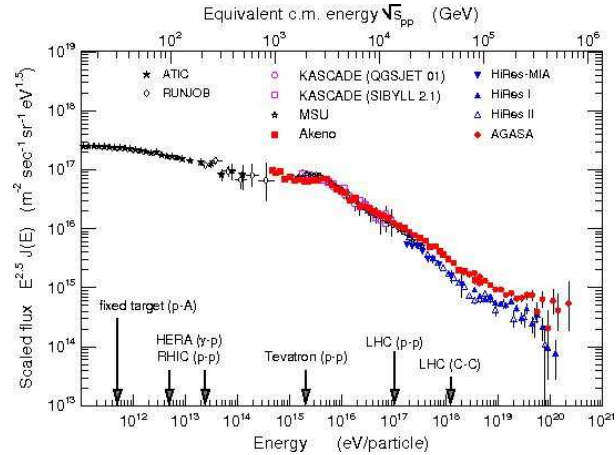}
  \caption{The cosmic proton spectrum with the "knee" between Tevatron and LHC energies.}
  \label{knee}
\end{figure}

The complete study of baryon spectra at LHC energies ($\sqrt(s)$=0.9 and 7. TeV) \cite {bylinkin} did not show important changes that may be responsible for the "knee" in CR proton spectra, see figure~\ref{knee}. 

It was observed that average $p_T$'s of baryons are slowly growing with energy. The average transverse momenta dependence on the mass of baryon (meson) at LHC energy is also considered up to the masses of charmed and beauty hadrons.

It is suggested in QGSM as well that the explanation of growing charge ratios of secondary antiparticle-to-particle spectra in cosmic rays can be done with the leading production asymmetry of baryon spectra toward the antibaryon spectra in the kinematical region of fragmentation of targets, which are mostly of positive matter. The procedure of spectra transfer from center-of-mass system at LHC p-p collisions into the laboratory system at cosmic ray interactions is given in the Appendix. 
 
\section{QGSM model for the production of baryons}
\label{sec-1}

Let us first describe the QGSM approach, which has been applied for recent studies of $\Lambda^0$. According to this approach baryon production can be parameterized in the following way:
\begin{equation}
\label{qgsm}
\frac{d\sigma}{p_T d p_T} = A_0\exp [-B_0\cdot (m_T - m_0)],
\end{equation}
where, $m_T=\sqrt(p_T^2+m0^2)$, $m_0$ is the mass of the produced hadron and $B_0$ is the slope parameter for the considered energy. In the early paper \cite{Lambda}, it was also shown that the value of the slope parameter $B_0$ becomes dependent on the collision energy. This approximation works well upto $p_T$= 6 GeV, that gives $\approx$ 85 percents of integral cross section. 

\section{Average transverse momenta of baryons at LHC}
\label{sec-2}
\begin{figure}[htpb]
\centering
\includegraphics[width=6cm]{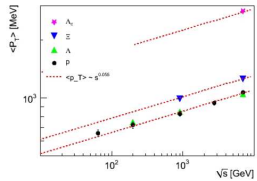}\\
\caption{Mean transverse momenta of charged baryons as a function of $\sqrt{s}$.}
\label{pt}
\end{figure}

Let us now discuss the mean transverse momenta of produced baryons and look at its dependence on the collision energy, $\sqrt{s}$. It is reasonable also to compare the values calculated for the proton spectra with other available data on baryon production: $\Lambda$, $\Xi$~\cite{CMSL} and $\Lambda_c$~\cite{LHCB} spectra.
Figure~\ref{pt} shows such dependence for the available experimental data. The steep rise of the mean transverse momenta $<p_T>$ with energy is seen in figure~\ref{pt}. Remarkably, this rise can be parameterized by the same power-like $s^{0.055}$ behavior in case of all the species of produced baryons. 

\begin{figure}[htpb]
\centering
\includegraphics[width=6cm]{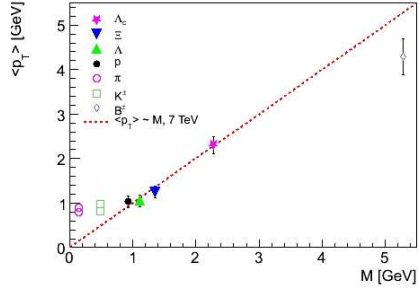}\\
\caption{Average transverse momenta of baryons as a function of their mass are presented at the energy 7 TeV.}
\label{m}
\end{figure}

Another interesting implication reveals in the comparison of the average transverse momenta of various produced baryons at the certain collision energy as a function of their masses, shown in figure~\ref{m}.
There a linear dependence between the mean transverse momenta $<p_T>$ and the baryon mass $M$ is observed. At $\sqrt{s} = 7$ TeV the average transverse momentum reaches the value of baryon mass, t.e. $<p_T> \sim M$. Further measurements at LHC Run-II should clarify whether or not the average transverse momentum expansion in the baryon production has a limit $<p_T> = M$.

It also would be interesting to compare $<p_T>$'s as a function of produced baryon mass $M$ with the mass dependence of the average transverse momenta that have been calculated from the description of charged meson production.

\section{Antiparticle-to-particle ratios as a result of leading baryon asymmetry}

\begin{figure}[htpb]
\centering
\includegraphics[width=6cm]{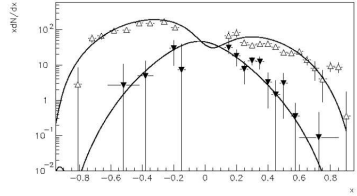}\\
\caption{The form of energy spectra of hyperons produced in the hyperon-proton collisions as compiled from two fixed target experiments.}
\label{fig:sigma} 
\end{figure}

The important feature on baryon production in proton-proton interactions is valuable asymmetry between antibaryon and baryon energy distributions in the kinematical regions of beam fragmentation. It is seen in the figure ~\ref{fig:sigma}, where the spectra of $\Lambda_c$ are described in QGSM for the entire kinematical $x_F$ range \cite{sigma}.

\begin{figure}[htpb]
\centering
\includegraphics[width=6cm]{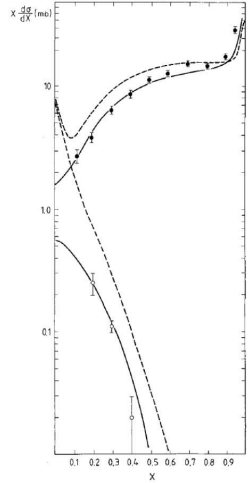}\\
\caption{Spectra of protons (black circles) and antiprotons (empty circles) in the low energy experiment and the spectra expected at high energy.}
\label{fig:proton_antiproton} 
\end{figure}

Larger asymmetry appears in the spectra of protons and antiprotons as it is seen in figure \ref{fig:proton_antiproton}.
What is also important, there is dip between the growing central part of distribution (the "table") and the stable proton fragmentation region.
Such whimsical form of c.m.s. spectra would give "knees" and "shoulders" at the end of spectra in laboratory system of cosmic ray interactions.

\begin{figure}[htpb]
\centering
\includegraphics[width=6cm]{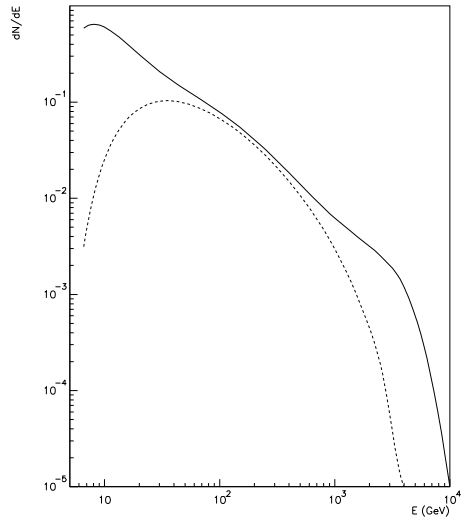}\\
\caption{The form of energy spectra of $\Lambda_c$ baryons produced in the proton-proton collisions in the laboratory system.}
\label{lambdalab} 
\end{figure}

In the laboratory system of cosmic ray interactions, these spectra are to be converted into the energy distributions like those shown in figure \ref{lambdalab}, where the valuable asymmetry takes place at the energies up to hundred GeV. The procedure of transformation of spectra at the pass from center-of-mass system into laboratory system of coordinates is presented in the Appendix. 

The growing ratio between antiparticle and particle spectra appears in the left side of spectra due to influence of positive proton target, as it was already learned from the early publication \cite {antiprotons}.

This suggestion makes the particle production at the sources of very high energy cosmic rays important. 

\section{Conclusions} 
Transverse momentum spectra of protons and antiprotons from RHIC ($\sqrt(s)$ = 62 and 200 GeV) and LHC experiments ($\sqrt(s)$= 0.9 and 7 TeV) have been described in the QGSM approach. This model seems working for the up-to-date collider energies, because spectra at low $p_T$ are still giving the main part of integral cross section. It seems that the enhancement of power-low contribution into the spectra at high $p_T$'s causes the change of low $p_T$ exponential slopes, so that the mean transverse momenta are growing with energy.
These average transverse momenta are also growing with mass if we are analyzing the dependences of spectra of different baryons on their masses. 
Nevertheless, all this changes in hadroproduction spectra cannot be dramatic enough to cause the "knee" in the primary proton spectra in cosmic rays at the energy in laboratory system correspondent to LHC energy.
Another aspect of application of QGSM approach is charge asymmetry of CR spectra. 
The growing antiparticle/particle ratios cannot be the result of acceleration of matter that is mostly positive around us.
Growing charge ratios of secondary particle spectra in CR may be generated by the baryon production in the CR interactions with positive matter targets.

\section{Appendix}
\begin{figure}[htpb]
\centering
\includegraphics[width=8cm,height=8cm]{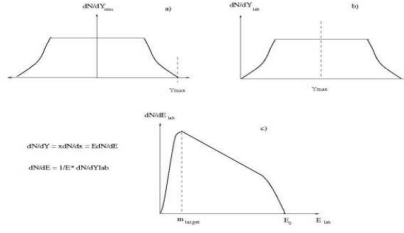}\\
\caption{The graphical procedure of recalculation of baryon dN/dY spectra measured in c.m. system at collider experiments into the spectra, dN/dE, as they are seen in laboratory system at cosmic ray measurements.}
\label{fig:appendix} 
\end{figure}

It was found in an earlier paper \cite {gamma} that the typical rapidity distributions of hadrons at collider experiments are easily convertible into dN/dE energy spectra in the laboratory system, where one beam particle becomes the target. This procedure is graphically illustrated in the figure~\ref{fig:appendix}. It was applied first time in \cite{gamma}.

As we know the rapidity distributions, dN/dY, at high energy proton collisions looks like the "table". Rapidity spectrum in laboratory system are reached by the shift of the "table" on the value of $Y_max$, so that all rapidity range is positive as in the fixed target collisions. Then our distribution should be expressed in energy variable x=E/$E_{collision}$: dN/dY=xdN/dx. It is clear that the energy spectrum in the laboratory system will be power like dN/dE $\approx$ 1/E.   

\section{Acknowledgements}
Author is sincerely thankful to Prof.K.Boreskov for numerous discussions and useful advises.

\end{document}